\input{aipcheck}

\documentclass[
    ,final            
  ]
  {aipproc}

\layoutstyle{6x9}

\def\alt{\mathrel{\mathpalette\gl@align<}}
\def\agt{\mathrel{\mathpalette\gl@align>}}
\def\gl@align#1#2{\lower.6ex\vbox{\baselineskip\z@skip\lineskip\z@
\ialign{$\m@th#1\hfil##\hfil$\crcr#2\crcr\sim\crcr}}} \makeatother

\def \gtsim    {\relax\ifmmode{\mathrel{\mathpalette\oversim >}}
                  \else{$\mathrel{\mathpalette\oversim >}$}\fi}
\def \ltsim    {\relax\ifmmode{\mathrel{\mathpalette\oversim <}}
                  \else{$\mathrel{\mathpalette\oversim <}$}\fi}
\def\oversim#1#2{\lower4pt\vbox{\baselineskip0pt \lineskip1.5pt
            \ialign{$\mathsurround=0pt#1\hfil##\hfil$\crcr#2\crcr\sim\crcr}}}

\newcommand{\gev}  {\mbox{${\rm GeV}$}}

\newcommand{\invfb}{\mbox{${\rm fb}^{-1}$}}

\newcommand{\faketau} {\mbox{$f_{j\rightarrow\tau}$}}
\newcommand{\pt}  {\mbox{$p_{\rm T}$}}
\newcommand{\ptvis}  {\mbox{$p_{\rm T}^{\rm vis}$}}
\newcommand{\et}  {\mbox{$E_{\rm T}$}}

\newcommand{\met} {\mbox{${E\!\!\!\!\!/_{\rm T}}$}}

\newcommand{\dM}{\mbox{$\Delta M$}}

\newcommand{\mtautaumax}{\mbox{$M_{\tau\tau}^{\rm end~point}$}}
\newcommand{\mtautauvis}{\mbox{$M_{\tau\tau}^{\rm vis}$}}
\newcommand{\mtautaupeak}{\mbox{$M_{\tau\tau}^{\rm peak}$}}

\newcommand{\azero}{\mbox{$A_{0}$}}
\newcommand{\tanb} {\mbox{$\tan\beta$}}

\newcommand{\mhalf}{\mbox{$m_{1/2}$}}

\newcommand{ \gluino}   {\mbox{$\tilde{g}$}}
\newcommand{ \squark}   {\mbox{$\tilde{q}$}}

\newcommand{ \stauone}  {\mbox{$\tilde{\tau}_{1}$}}

\newcommand{ \stau}     {\mbox{$\tilde{\tau}$}}

\newcommand{ \schionezero }{\mbox{$\tilde{\chi}_{1}^{0}$}}
\newcommand{ \schitwozero }{\mbox{$\tilde{\chi}_{2}^{0}$}}

\newcommand{ \isajet }    {{\tt ISAJET}}
\def \etal     {\relax\ifmmode{et \; al.}\else{$et \; al.$}\fi}

\def \calR     {\relax\ifmmode{{\cal R}}\else{${\cal R}$}\fi}
\begin{document}

\title{Detection of SUSY Signals in Stau Neutralino Co-annihilation Region at the LHC}

\classification{11.30.Pb, 12.60.Jv, 14.80.Ly}

\keywords      {Neutralino dark matter, Stau neutralino co-annihilation, LHC}

\author{R. Arnowitt}{
  address={Department of Physics, Texas A\&M University, USA}
}
\author{A. Aurisano}{
  address={Department of Physics, Texas A\&M University, USA}
}
\author{B. Dutta}{
  address={Department of Physics, Texas A\&M University, USA}
}

\author{T. Kamon}{
  address={Department of Physics, Texas A\&M University, USA}
}

\author{N. Kolev}{
  address={Department of Physics, University of Regina, Canada}
}

\author{P. Simeon}{
  address={Department of Physics, Texas A\&M University, USA}
}

\author{D. Toback}{
  address={Department of Physics, Texas A\&M University, USA}
}

\author{P. Wagner}{
  address={Department of Physics, Texas A\&M University, USA}
}

\begin{abstract}
We study the prospects of detecting the signal in
the stau neutralino
($\stauone$-$\schionezero$) co-annihilation region at the LHC using tau
($\tau$) leptons. 
The co-annihilation signal is characterized by
the $\stauone$ and $\schionezero$  mass difference ($\dM$) to be 5-15 GeV
to be consistent with the WMAP measurement of the cold dark matter relic density 
as well as all other experimental bounds 
within the minimal supergravity model.
Focusing on $\tau$'s from 
$\schitwozero \to \tau \stau \to \tau \tau \schionezero$ decays
in \gluino\ and \squark\ production, 
we consider inclusive $\met$+jet+3$\tau$ production, 
with two $\tau$'s above a high \et\ threshold and a third $\tau$ above a lower threshold.  
Two observables, the number of opposite-signed $\tau$ pairs minus 
the number of like-signed $\tau$ pairs and the peak position of 
the di-tau invariant mass distribution, allow for 
the simultaneous determination of \dM\ and $M_{\gluino}$.  
For  \dM\ = 9~GeV and $M_{\gluino}$ = 850~GeV with 30 \invfb\ of data, 
we can measure \dM\ to 15\% and $M_{\gluino}$ to 6\%.

\end{abstract}

\maketitle


\section{Introduction}

Supersymmetry (SUSY) with
$R$-parity invariance automatically gives rise to a candidate,
the lightest neutralino ($\schionezero$),
for the astronomically observed cold dark matter (CDM),
deeply linking particle physics with cosmology.
If SUSY is correct, the $\schionezero$ particles
will copiously be produced at the LHC. 
With the recent WMAP measurement of the CDM relic density~\cite{WMAP}
along with
other experimental results,
the mostly allowed parameter space within mSUGRA~\cite{mSUGRA,mSUGRA2}
is the stau-neutralino
($\stauone$-$\schionezero$) co-annihilation region~\cite{CA}.
We investigate the accelerator phenomena at the LHC.

\section{SUSY Signal in the  Co-annihilation Region}

The co-annihilation signal is characterized by a narrow mass difference ($\dM$)
between $\stauone$ and $\schionezero$ of  about 5-15 GeV~\cite{2tauAnalysis}.
 Thus if this striking near degeneracy between
$\stauone$ and $\schionezero$ is observed at the LHC, it would be a
strong indirect indication that the $\schionezero$ is the
astronomical CDM particle. 

\begin{figure}
\rotatebox{-90}{\includegraphics[height=.4\textheight]{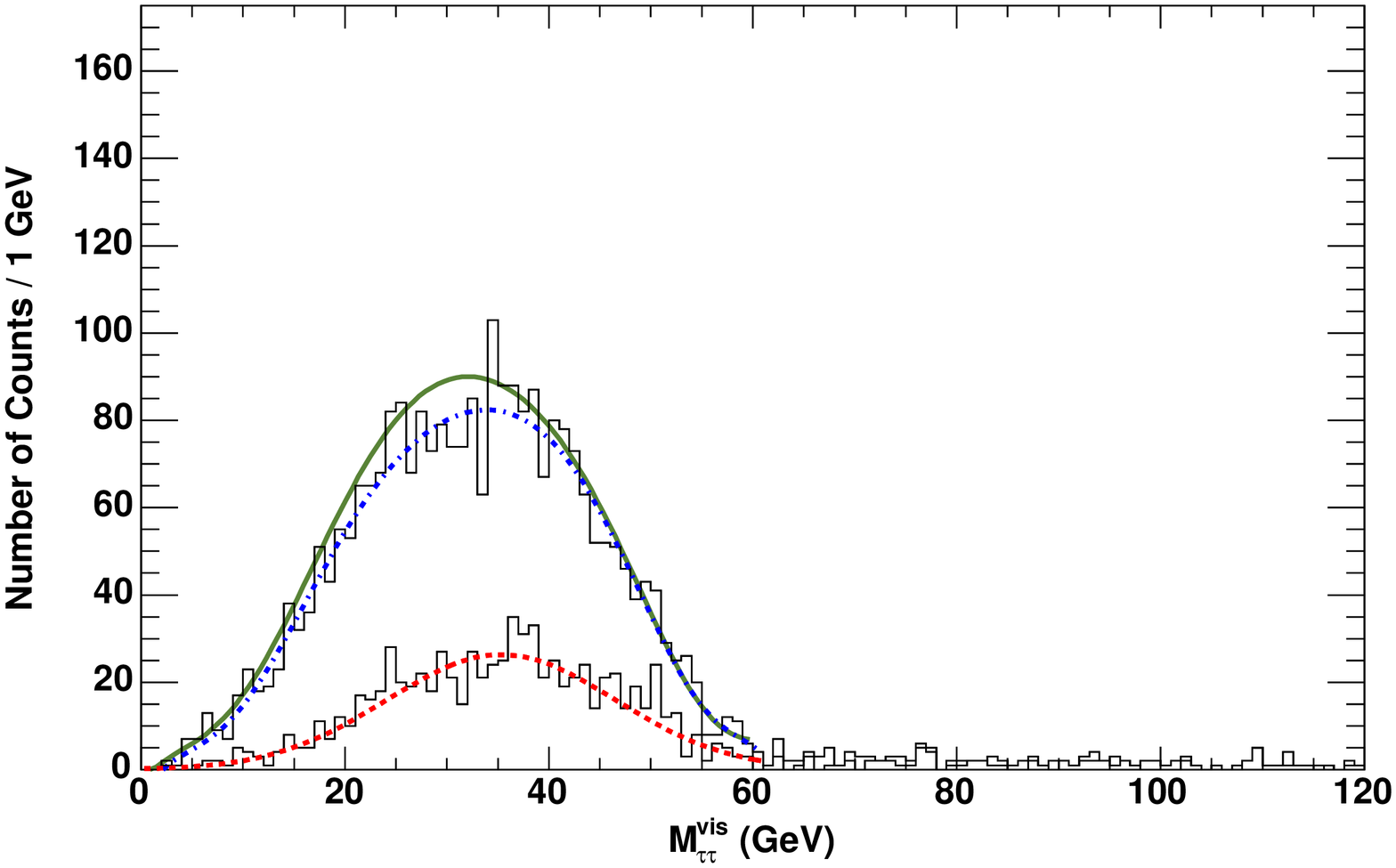}}  
  \caption{Visible di-tau mass distributions at a generator level for  
	hadronically decaying $\tau$ leptons from
	$\schitwozero \to \stauone \tau \to \schionezero \tau \tau$
	in mSUGRA events ($\dM$ = 5.7 GeV; the
$\gluino, \schitwozero, \schionezero$, and $\stauone$ masses are
831, 264.4, 137.4, and 143.1 GeV, respectively).
	Three curves (solid, dash-dot, and dotted) are the distributions for
	$\tau$'s selected with
	$\ptvis\ >$ (20, 20), (40, 20), and (40, 40) GeV, respectively. 
	Reconstructing $\tau$'s
	with $\ptvis\ > 20\ \gev$ is vital.  }
\label{fig:visDitauMass_mSUGRA}
\end{figure}

Focusing on $\tau$'s from 
$\schitwozero \to \tau \stau \to \tau \tau \schionezero$ decays
in \gluino\ and \squark\ production, 
we first examine the visible mass (\mtautauvis) distributions
with three choices of $\ptvis$ threshold values 
for events in an mSUGRA co-annihilation region
($M_{\gluino}$ = 831 GeV and $\dM$ = 5.6 GeV
for $\tanb$ = 40, $\mu > 0$, and $\azero\ = 0$) 
using {\tt ISAJET}~7.69~\cite{ISAJET}.
Even with such a small mass difference, 
the  lower energy $\tau$ is boosted in the cascade decay of the heavy
squark and gluino making it potentially viable
with $\ptvis\ \gtsim\ 20\ \gev$ as shown in 
Fig.~\ref{fig:visDitauMass_mSUGRA}.
We also note the end point of the mass distribution can be
inferred from the figure, which is 62 GeV by
\begin{eqnarray} 
\label{eq:mtautaumax}
\mtautaumax  & = &
M_{\schitwozero}
\sqrt{1 - M_{\stauone}^2 / M_{\schitwozero}^2 }
\sqrt{1 - M_{\schionezero}^2 / M_{\stauone}^2 }~.
\nonumber
\end{eqnarray}
From here on, we assume that both the ATLAS and CMS detectors can
reconstruct and identify $\tau$'s with  $\ptvis > 20\ \gev$ at
an efficiency of $\epsilon = 50\%$. 

We consider two experimental final states:
(a) $\met$ + $\geq$2~jet + $\geq 2~\tau$ events
($2\tau$ analysis)  \cite{2tauAnalysis};
(b)  $\met$ + $\geq$1~jet + $\geq 3~\tau$ events 
($3\tau$ analysis) \cite{3tauAnalysis}.
The signal in $3\tau$ analysis occurs at a reduced rate, but with lower background.
We calculate \mtautauvis\ for every pair of
$\tau$'s in the event  and categorize as opposite sign (OS) or like
sign (LS). 
The mass distribution for LS pairs is subtracted from the
distribution for OS pairs to extract $\schitwozero$ decays on a
statistical basis \cite{HP_large_dM}.  
We note both final states will be triggered by
requiring large \et\ jet(s) and large \met\, which will
be available at the ATLAS and the CMS experiments. 

Since two analyses use the same OS$-$LS technique, 
we only describe $3\tau$ analysis in this paper, where
the number of OS$-$LS $\tau$ pairs ($N_{\rm OS-LS}$)
and the peak position of the di-tau invariant mass distribution
(\mtautaupeak)
are used for the simultaneous determination of  
$\dM$ and $M_{\gluino}$. 

\begin{figure}
  \includegraphics[height=.3\textheight]{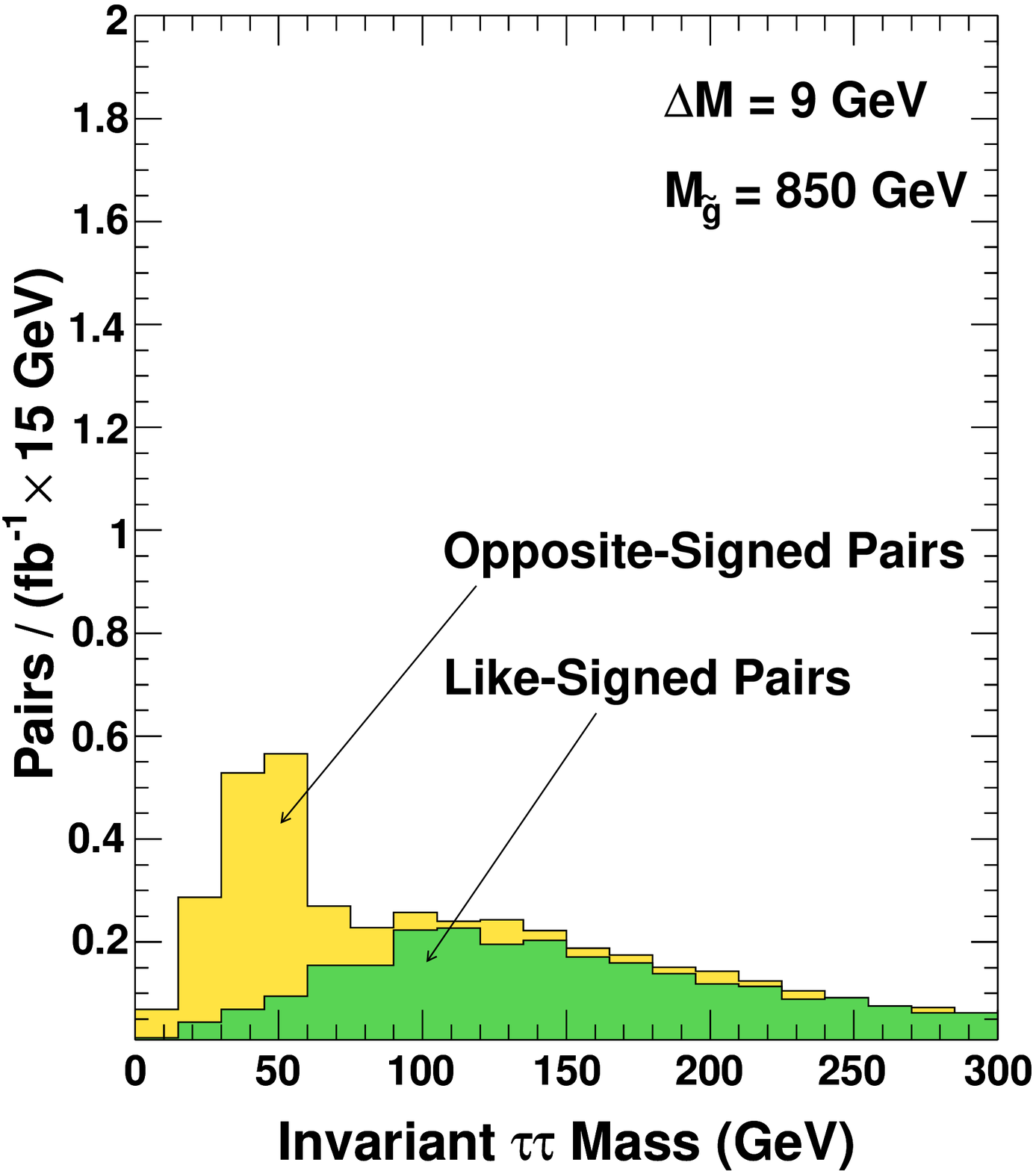}
  \includegraphics[height=.3\textheight]{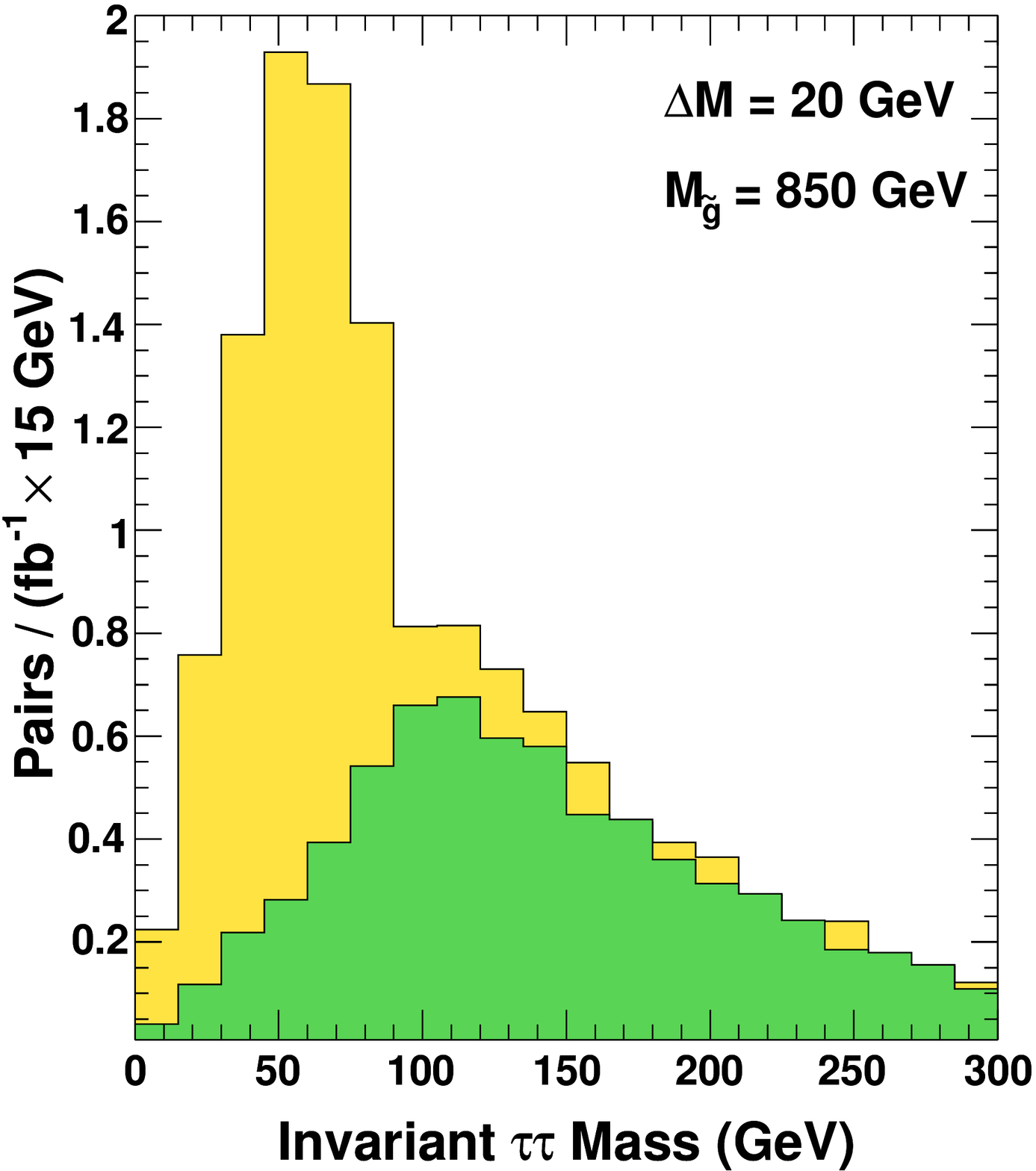}
\caption{Visible di-tau invariant mass (\mtautauvis) distributions
in the co-annihilation region ($\dM = 9\ \gev$ and $20\ \gev$) 
in the 3tau analysis with $\epsilon = 50\%$ and \faketau\ = 1\%.
 }
\label{fig:ditau_mass_shape_in_3tauAnalysis}
\end{figure}

We use
\isajet~7.64~\cite{ISAJET}
and PGS~\cite{PGS}, and
select events with 
at least one jet with $E_{\rm T} >$100 GeV and $\met >$ 100 GeV,
followed by requiring at least two identified
$\tau$'s with $\ptvis > 40\ \gev$, 
additional one with $\ptvis > 20\ \gev$,  and
$E_{\rm T}^{~{\rm jet 1}} + \met > 400\ \gev$. 
Figure~\ref{fig:ditau_mass_shape_in_3tauAnalysis} 
is the mass distributions in the 3tau analysis, where we have assumed $\epsilon$ = 50\% and a
probability that a jet is misidentified 
as $\tau$ (\faketau) to be 1\%. 
We see that the non-$\tilde\chi^0_2$ OS  pairs are nicely
canceled with the wrong LS combination pairs and that the OS$-$LS
distribution is well fit to a Gaussian.
We note that \mtautaupeak\
increase as the $\dM$ increases.

Within mSUGRA models, \mtautaupeak\ changes with $\dM$
because the $\gluino$, $\schitwozero$
and $\schionezero$ masses are related.
In the co-annihilation region, 
the squark masses largely depends on $\mhalf$,
so that the production cross section ($\sigma$) is mainly
as a function of $M_{\gluino}$. The event acceptance ($A$)
depends on  $\dM$ and $M_{\gluino}$.
We parameterize
$\mtautaupeak\ = F(\Delta M, M_{\gluino})$ and
$N_{\rm OS-LS}
= \sigma(M_{\gluino}) \cdot A(\dM, M_{\gluino}) \cdot \mathcal{L}$.
Here $\mathcal{L}$ is a luminosity.

\begin{figure}
  \includegraphics[height=.3\textheight]{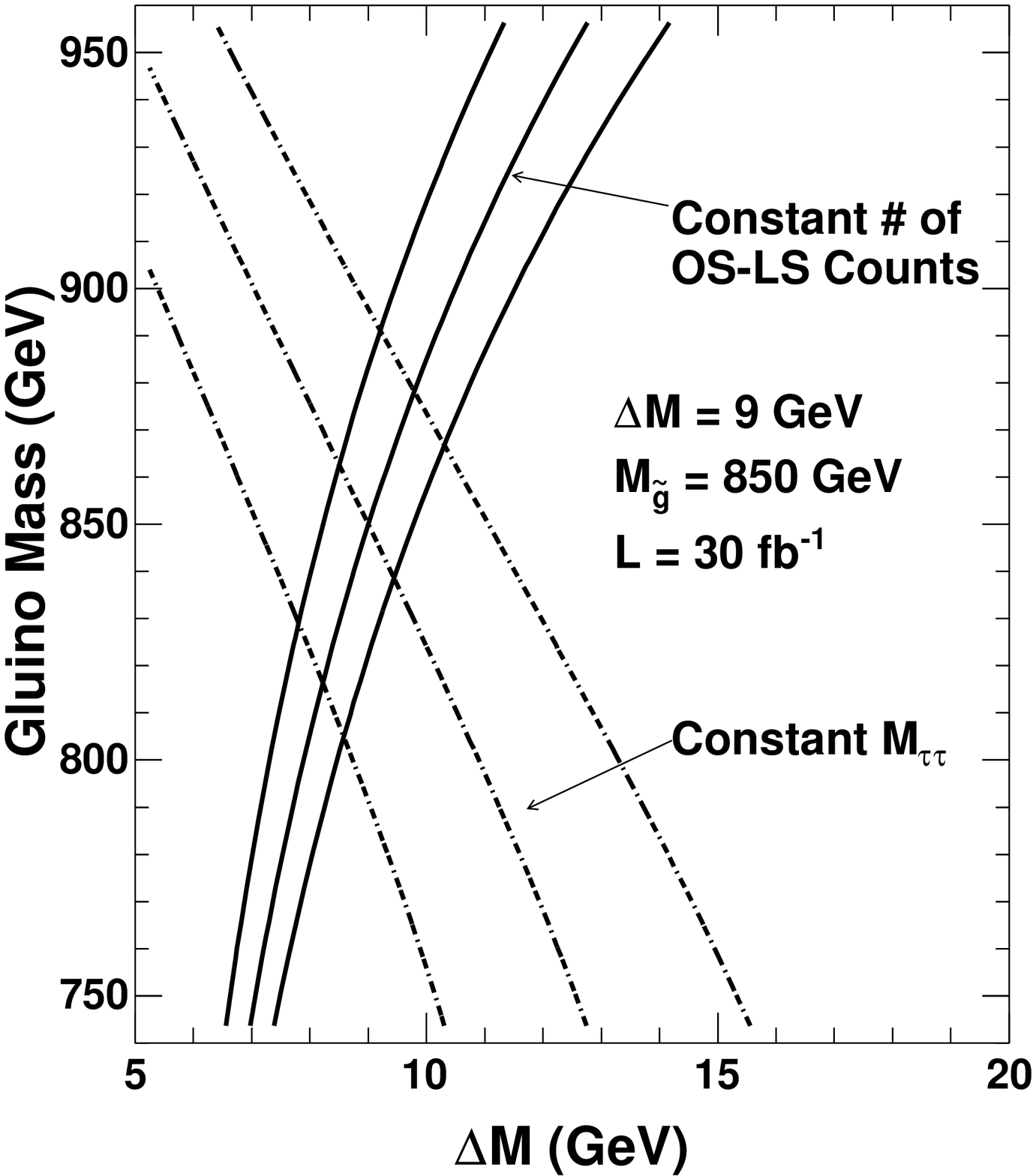}
\caption{Contours of constant $N_{\rm OS-LS}$  and \mtautaupeak\ for 
\dM = 9~GeV, $M_{\gluino}$ = 850~GeV, and $\mathcal{L}$ = 30~fb$^{-1}$.
 The middle lines are the central values while the outer lines show the 1$\sigma$ 
 uncertainty on the measurements.  
 The region defined by the outer four lines indicates the 1$\sigma$ region 
 for the \dM and $M_{\gluino}$ measurements.} 
\label{fig:mg_dm_in_3tauAnalysis}
\end{figure}

A correlation between the two functions allow us
to measure $M_{\gluino}$ and $\dM$ 
using \mtautaupeak\ and  $N_{\rm OS-LS}$.
Figure~\ref{fig:mg_dm_in_3tauAnalysis} shows  
the contours of constant $N_{\rm OS-LS}$  and \mtautaupeak\ for 
\dM = 9~GeV, $M_{\gluino}$ = 850~GeV,
and $\mathcal{L}$ =  30 \invfb. 
We find that for $\mathcal{L}$ = 30~fb$^{-1}$, 
we can measure  $\dM$ to $\sim$15\% and $M_{\gluino}$ to $\sim$6\%.  
It is important to note, however, that our determination 
of  $M_{\gluino}$ is not a direct measurement, 
but a determination of the SUSY mass scale of the model.  
It will need to be compared to a direct  $M_{\gluino}$ measurement, 
assuming one is available.  If the two results were consistent, 
it  would be a consistency check of the gaugino universality 
of the mSUGRA model and that we are indeed in the
 co-annihilation region.
Further, we expect that this analysis and the 2$\tau$ analysis 
could be used to complement each other 
in the establishment of a co-annihilation signal at the LHC, 
and perhaps be combined to produce a more accurate measurement.  

\section{Conclusion}

We have demonstrated
that if LHC experiments reconstruct/identify $\tau$'s  with $\pt >$
20 GeV with an efficiency in the $50\%$ range, we could establish the
signal in this  co-annihilation region  
by detecting $\schitwozero \rightarrow \tau
\stauone \rightarrow \tau\tau\schionezero$.
For our mSUGRA reference point of \dM\ = 9~GeV and $M_{\gluino}$ = 850~GeV,
we can measure \dM\ to $15\%$ and $M_{\gluino}$ to $6\%$ 
with 30~\invfb\ by
simultaneously measuring the number of OS$-$LS di-tau pairs 
and the peak position of OS$-$LS di-tau mass distribution in
the inclusive $\met + \mathrm{jet} + 3\tau $ final state.
A $15\%$ measurement of \dM\ would generally be sufficient 
to determine if the signal is consistent with co-annihilation, 
and therefore, with the $\schionezero$ being the dark matter particle. 

\begin{theacknowledgments}

This work was supported in part by DOE Grant
DE-FG02-95ER40917 and NSF grant DMS 0216275.

\end{theacknowledgments}


\end{document}